# Mediation of hydrogen-bond coupling interactions by programmable heating and salting

Xi Zhang,[1,2,a] Yongli Huang,[3,a] Zengsheng Ma[3], Yichun Zhou[3,] and Chang Q Sun[1,2,3*]

[1] NOVITAS, School of Electrical and Electronic Engineering, Nanyang Technological University, Singapore 639798

[3] Center for Coordination Bond and Electronic Engineering, College of Materials Science and Engineering, China Jiliang University, Hangzhou 310018, China

[3] Key Laboratory of Low-dimensional Materials and Application Technology (Ministry of Education) and Faculty of Materials, Optoelectronics and Physics, Xiangtan University, Xiangtan, 411105, China

Ecqsun@ntu.edu.sg

[a] X.Z. and Y.H. contribute equally.

Abstract

We show that programmable heating and salting share the same effect on the frequency shift of the O:H and the H-O stretching phonons of the O:H-O hydrogen bond, which revealed that both heating and salting lengthens and softens the O:H bond and shortens and stiffens the H-O bond due to the weakening of the Coulomb repulsion between electron pairs of adjacent oxygen atoms. Understanding provides possible mechanism for the Hofmeister series and the detergent effect on cloth cleaning.

The behavior of aqueous ions has profound effects on biological molecules such as proteins and DNA and, thus, implications for health and disease. The ions added in the form of salts or buffer reagents to the protein solution are crucial for maintaining protein stability. Different ions are better or worse at preventing aggregation and self-association. At the core of the problem lies a phenomenon discovered in the late 1800s by the Czech chemist Franz Hofmeister. He discovered that certain aqueous ions follow a peculiar order in their increasing or decreasing ability to precipitate egg whites in solution. Anions in particular, such as $SO_4^{2-}$, $Cl^-$, and $SCN^-$, follow a seemingly arbitrary sequence: In this order, they increasingly can denature and dissolve proteins, and have increasing or decreasing effects on many other solution properties, such as surface tension [1].

Randall and Failey [2-4] realized that the efficiency of some common cations as salting-out agents follows the orders:

For cations: $Na^+ > K^+ > Li^+ > Ba^{2+} > Rb^+ > Ca^{2+} > Ni^{2+} > Co^{2+} > Mg^{2+} > Fe^{2+} > Zn^{2+} > Cs^+ > Mn^{2+} > Al3^+ > Fe^{3+}, Cr^{3+} > NH4^+ > H^+$

For anions: $OH^- > SO_4^{2-}, CO_3^{2-} > ClO_4^- > BrO_3^- > Cl^- > CH_3COO^- > IO_3^-, IO_4^- > Br^-, I^- > NO_3^-$

The effect of the addition of salts on solutions of nonelectrolytes is very complex, due to the different types of intermolecular interactions that involve the ions, the solvent, and the solute molecules [5].

There are two explanations of the mysterious Hofmeister series. One was that ions produced long-range effects on the structure of water, leading to changes in water's ability to let proteins fall out of, or stay dissolved in, a solution. That idea has largely been discarded. The current view is that Hofmeister effects stem largely from the varying abilities of different salt ions to replace water at nonpolar molecular or macroscopic surfaces. But no theoretical framework can yet predict these actions [5]. Understanding the impact of ions on the properties of aqueous solutions and how these modified properties influence chemical and conformational dynamics remains an important and elusive objective of physical chemistry research. Sweet solutions can be obtained upon heating mixtures of simple carbohydrates, urea and inorganic salts to moderate temperatures, to give new chiral media for organic reactions [6]. Salt can assist the dissociating snow to the public traffic conditions upon a heavy snowing in cold zones. Therefore, sugar and salt addition could lower the dissociation energy and hence anti-icing takes place. The solubility of sugar increases with temperature but drops with pressure [7].

Using ultrafast 2DIR vibrational spectroscopy and MD simulations, Park and coworkers have found [8] that NaBr addition to 5% HOD in $H_2O$ shifts the O-D stretching frequency from 2509 to 2539 cm$^{-1}$ depending on the relative number (8, 16, 32) of $H_2O$ molecules per Br$^-$ ion. In aqueous NaBr solutions, the hydrogen bond network of water is partially or completely disrupted depending on the concentration of NaBr salt. The HOD molecules are hydrogen-bonded to ions as HOD–Br$^-$, DOH–Br$^-$, and HDO–Na$^+$ in the hydration shells around the ions. Salt ions such as (NaCl [9], NaBe [8], LiCl [10], NaClO$_4$ and Mg(ClO$_4$)$_2$ [11-13]), induced blue shift of the hydroxyl group (-OH or -OD) have been detected by the same group. Aqueous LiCl performs the same [10] to drop the supercooling temperature from 248 K to 190 K by adding $H_2O$/LiCl ratio from 100 to 6.7. The vibration frequency also changes with the PH value of organic liquid such as $NH_4H_2PO_4$ [14]. These features share the same attribute of the heating effect on H-O vibration [9, 15].

In order to exam our expectations on the salt mediation of the Coulomb coupling in the H-bond, we conducted FTIR measurement of liquid water with different concentration of NaCl% in comparison of clean water at changed temperatures. We found that the salting has the same effect on the frequency shift of the $\omega_H$ and $\omega_L$ phonons. Figure 1 shows the residual FTIR $\omega_{B1}$ (~550 cm$^{-1}$ corresponding to the O:H-O being) and $\omega_H$ (3200~ 3450 cm$^{-1}$) spectra of water upon being heated and NaCl salted. The salted $\omega_H$ spectra are the same as reported in [9, 17]. Heating and salting effect the same on stiffening the stiffer phonon and softening the softer phonons. The identical effect of salt and heat indicates that heating and salting modulates the repulsion between electron pairs of oxygen atoms in a common mechanism of O:H-O relaxation dynamics.

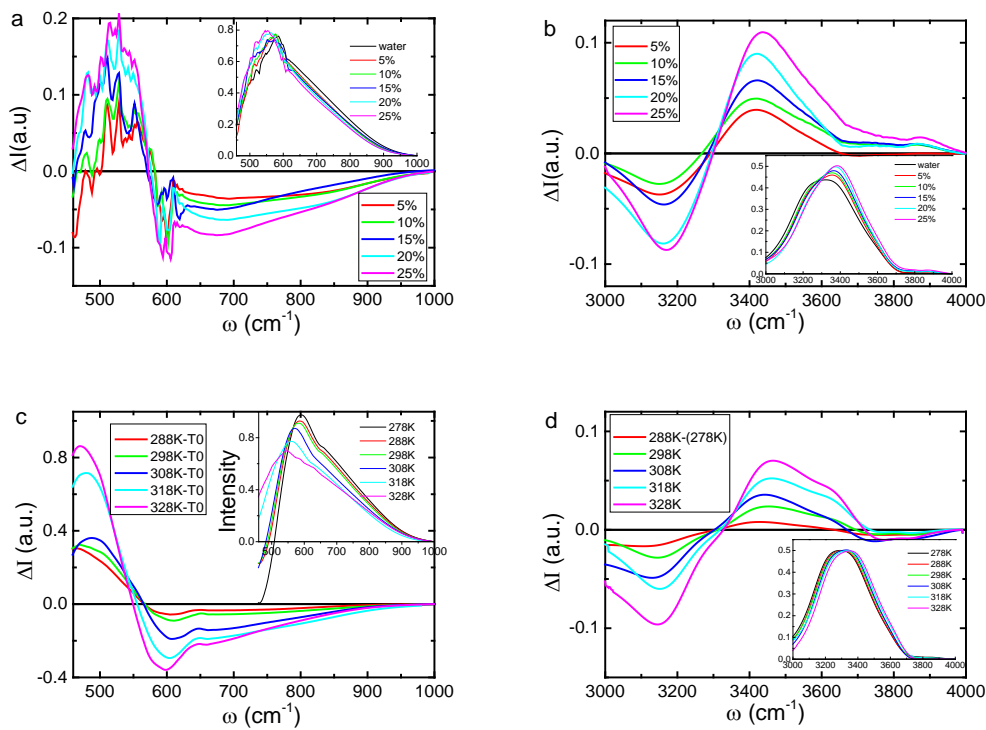

Figure 1 FTIR spectra of water as a function of salinity (mass% NaCl) and heating. Insets show the Residual spectra of treated water with respect to the referential spectrum collected at 278 K without salting. Heating and salting effect the same on stiffening $\omega_H$ phonons and softening the $\omega_L$ [16].

As we found recently [9, 10, 16, 19, 20], heating lengthens and softens the O:H noncovalent bond and meanwhile shortens and stiffens the H-O covalent bond, results in the lengthening of the O---O distance. The lengthening of O---O weakens the Coulomb repulsion. The identical effect of heating and salting indicates that replacement of O ions with anions or cations has the same effect on the Coulomb repulsion between the ions.

One can imagine that heating lengthens the $d_{OO}$ and weakens the Coulomb repulsion. Replacing an $O^{2\delta-}$ ion with the Cl$^-$ and the Na$^+$ ion, the repulsion between the guested ions and the oxygen could be modulated – which could be the possible mechanism for the Hofmeister series [1, 5, 18]– cations and anions not only decreases the ability to augment water surface tension but also increase the ability to dissolve proteins. Soaps, sugars, and detergents may functionalize the same as heating and salting in washing and cleaning. The replacement of $O^{\delta-}$ ions with ions of salts, acids, or sugars and heating could weaken the Coulomb repulsion between the charged ions. The reduced repulsion restores the O in the H-O part to relax towards coordination origin and therefore shortens and stiffens the H-O bond [19]. According to the present understanding, weakening the repulsion will shorten the H-O bond and deepen its potential to raise the melting point and stiffening the high-frequency phonons.

Further molecular dynamics calculations of doped ice at 170 K suggested that the phonon frequencies are more sensitive to the cations then to the anions. Therefore, modulation of Coulomb repulsion (charge density and molecular separation) results in the phonon relaxation.

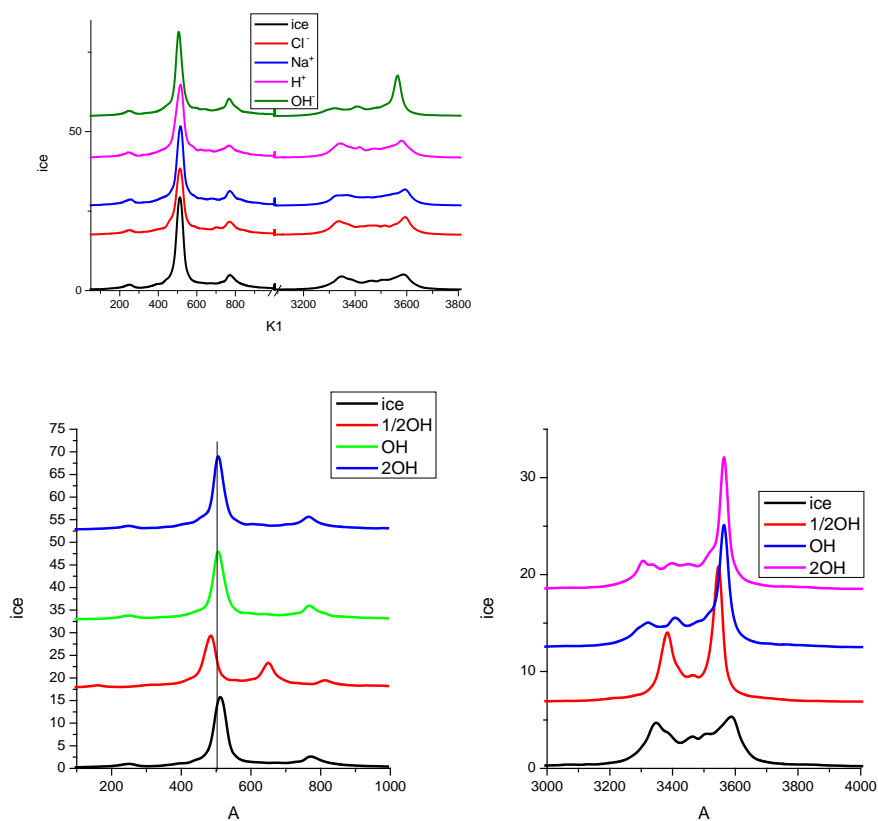

Figure 2 Na$^+$, Cl-, and H$^+$ and OH- cconcentration dependence on the frequency shift in 96-mol ice, under 170 K, in NPT ensemble for 100ps.

Aqueous ionic solutions lubricate the chemical machinery of the environment and life. Understanding the impact of ions on the properties of aqueous solutions and how these modified properties influence chemical and conformational dynamics remains an important and elusive objective of physical chemistry research. Sweet solutions can be obtained upon heating mixtures of simple carbohydrates, urea and inorganic salts to moderate temperatures, to give new chiral media for organic reactions [6]. Salt can assist the dissociating snow to the public traffic conditions upon a heavy snowing in cold zones. Therefore, sugar and salt addition could lower the dissociation energy and hence anti-icing takes place. It has been found that solubility of sugar increases with temperature but drops with pressure [7].

Using ultrafast 2DIR vibrational spectroscopy and MD simulations, Park and coworkers have found [8] that NaBr addition to 5% HOD in H$_2$O shifts the O-D stretching frequency from 2509 to 2539 cm$^{-1}$ depending on the relative number (8, 16,

32) of $H_2O$ molecules per $Br^-$ ion. In aqueous NaBr solutions, the hydrogen bond network of water is partially or completely disrupted depending on the concentration of NaBr salt. The HOD molecules are hydrogen-bonded to ions as $HOD–Br^-$, $DOH–Br^-$, and $HDO–Na^+$ in the hydration shells around the ions. Salt ions such as $NaClO_4$, $Mg(ClO_4)_2$, induced blue shift of the hydroxyl group (-OH or -OD) have been detected by the same group [11-13]. Aqueous LiCl performs the same [10] to drop the supercooling temperature from 248 K to 190 K by adding $H_2O$/LiCl ratio from 100 to 6.7.

Figure 1 shows the Raman data of water with and without salt [9, 17]. An addition of NaCl also shifts the $\omega_H$ to higher frequencies and enhances the thermal stiffening of $\omega_H$. The similar effect of salt and heat may indicate the commonly lengthening and softening of the non-bond and the $\omega_L$. This finding evidences that the dissociation energy is related to the non-bond interaction only. Both heating and salting may help in understanding the de-icing and anti-icing as a consequence of non-bond weakening. Soaps and detergents may functionalize the same as heating in washing and cleaning. [20, 21]